\documentclass[12pt,a4paper,oneside]{article}

\usepackage{amsmath}
\usepackage[english]{babel}

\begin{document}

\begin{figure}[!t]
  \begin{flushright}
    INFN-FE-04-98 \\
    hep-ph/9804464
  \end{flushright}
\end{figure}

\title{Once more on the bound on $\nu_\tau$ magnetic moment from L3 data}
\author{M.~Maltoni$^{a,b}$ and M.~I.~Vysotsky$^{b,c}$}
\date{}

\maketitle

\small\noindent
${}^a$ Dipartimento di Fisica, Universit\`a di Ferrara, I-44100,
Ferrara (Italy) \\
${}^b$ INFN, Sezione di Ferrara, I-44100, Ferrara (Italy) \\
${}^c$ ITEP, Moscow, Russia

\bigskip

\begin{abstract}

We show that recently announced strong bound on $\mu_{\nu_\tau}$ can not be
justified, and confirm original L3 result.

\end{abstract}

\section{Introduction}

Experimental bounds on the magnetic moment of the $\tau$-neutrino is much
weaker than bounds on magnetic moments of electron and muon neutrinos:
\begin{gather}
  \mu_{\nu_e} < 1.8 \times 10^{-10} \, \mu_B; \\
  \mu_{\nu_\mu} < 7.4 \times 10^{-10} \, \mu_B; \\
  \mu_{\nu_\tau} < 4 \times 10^{-6} \, \mu_B; \label{eq02} \\
  \mu_{\nu_\tau} < 5.4 \times 10^{-7} \, \mu_B; \label{eq05}
\end{gather}
where bound~(\ref{eq02}) comes from analysis of the $e^+ e^-$ annihilation
to $\gamma + \textrm{nothing}$ at low energies~\cite{Grotch88}, and
bound~(\ref{eq05}) comes from beam-dump experiment~\cite{Cooper92}. New
bound was obtained recently by L3 collaboration from analysis of the $e^+
e^-$ annihilation at the $Z$ resonance. Search for energetic single photon
production in $Z$ decays leads to the following bound~\cite{L3}:
\begin{equation}
  \mu_{\nu_\tau} < 3.3 \times 10^{-6} \, \mu_B. \label{eq10}
\end{equation}
However, in paper~\cite{Maya98} new analysis of L3 data was performed, and
much more stringent bound was announced:
\begin{equation}
  \mu_{\nu_\tau} < 1.14 \times 10^{-9} \, \mu_B,
\end{equation}
which(if correct) will put bound on $\mu_{\nu_\tau}$ close to that for
electron and muon neutrinos. Trying to reproduce result~\cite{Maya98} we
fail and confirm bound~(\ref{eq10}) obtained originally by L3 collaboration.

\section{Discussion}

In paper~\cite{Maya98} the following relation was used:
\begin{gather}
  \mathcal{L}_I = -\frac12 \mu_\nu \, F_{\mu\nu} \, \bar\nu \Sigma^{\mu\nu} 
  \nu - Z_\mu \, \bar\nu \gamma^\mu \left( g_V - \gamma_5 \, g_A \right) \nu, 
  \label{eq15} \\
  \mu_\nu \equiv \frac{\epsilon_6}{\sqrt2 v}, \qquad 
  g_V = g_A \equiv \frac g{4 c_w} = \frac e{4 c_w s_w}, \label{eq16}
\end{gather}
where $v$ is the Higgs boson vacuum expectation value, $v \approx 246 \, GeV$.
For the partial decay width $Z \to \nu \bar\nu \gamma$
from~(\ref{eq15}) it was obtained:
\begin{equation} \label{eq17}
  \frac{d\Gamma^{(a)}}{dx} = 
  \frac{\epsilon_6^2 \left( g_V^2 + g_A^2 \right) M_Z^3}  
  {72 \pi^3 v^2} x \left( 3 \left( 1-2x \right) + x^2 \right).
\end{equation}
In paper~\cite{Maya98}, an additional effective $Z \nu \bar\nu \gamma$ vertex
was introduced, leading to the following partial decay width $Z \to \nu
\bar\nu \gamma$ (there is no interference with $\mu_\nu$ induced decay):
\begin{equation} \label{eq18}
  \frac{d\Gamma^{(b)}}{dx} = 
  \frac{\epsilon_8^2 M_Z^5}{18 \pi^3 v^4} x^3 \left( 1-x \right).
\end{equation}
Following the strategy outlined by Maya et al., we perform integration
of these equations to obtain the width for $Z \to \nu \bar\nu \gamma$
decay. The integration must be performed for $x$ ranging from
$\frac 14$ (the L3 collaboration required the photon energy to be
greater than half the beam energy~\cite{L3}) and $\frac 12$ (the maximum
value for the photon energy is reached when the photon direction is opposite
to the direction of the two neutrinos, and equals $\frac{M_Z}2$). We find:
\begin{gather}
  \Gamma^{(a)} = 
  \frac{\epsilon_6^2 \left( g_V^2 + g_A^2 \right) M_Z^3}  
  {72 \pi^3 v^2} I,
  \\  
  I = \int_{\frac 14}^{\frac 12} 
  x \left( 3 \left( 1-2x \right) + x^2 \right) dx = \frac{79}{1024} 
  \approx 0.077
\end{gather}
from eq.~(\ref{eq17}), and
\begin{gather}
  \Gamma^{(b)} = 
  \frac{\epsilon_8^2 M_Z^5}{18 \pi^3 v^4} J,
  \\
  J= \int_{\frac 14}^{\frac 12}
  x^3 \left( 1-x \right) dx = \frac{11}{1280} \approx 0.0086
\end{gather}
from eq.~(\ref{eq18}). We now remind that the L3 collaboration data sample
corresponds to $N_{Z \to \textrm{had}} = 3.3 \times 10^6$ hadronic $Z$
decays. The $Z \to \textrm{hadrons}$ decay width can easily be evaluated
neglecting strong interactions between quarks in the final state and
considering $Z \to q \bar q$ decay at tree level:\footnote{At the $Z$
resonance, 1-loop corrections give a contribution of order $\frac
{\alpha_s}{\pi} \approx 4\%$; we use parton model calculation instead of
experimentally measured $\Gamma_{Z \to \textrm{had}}$ width to get simple
analytical expressions~(\ref{eq25}) and~(\ref{eq30}).}:
\begin{gather}
  \Gamma_{Z \to q \bar q} =
  \frac{M_Z^3}{48 \pi v^2} \left[
  \left( C_V^q \right)^2 + \left( C_A^q \right)^2
  \right], \label{eq20}
  \\
  C_A^u = 1, \quad C_V^u = 1 - \frac 83 s_w^2,
  \\
  C_A^d = 1, \quad C_V^d = 1 - \frac 43 s_w^2.
\end{gather}
Only five quark flavours ($u, c$ and $d, s, b$) give contribution
to~(\ref{eq20}), so we have:
\begin{gather}
  C = 3 \left\{ 
    2 \left[ \left( C_V^u \right)^2 + \left( C_A^u \right)^2 \right] +
    3 \left[ \left( C_V^d \right)^2 + \left( C_A^d \right)^2 \right] \right\} 
  \approx 20.4,
  \\
  \Gamma_{Z \to \textrm{had}} =
  \frac{M_Z^3}{48 \pi v^2} C \approx 1.69 \; GeV,
\end{gather}
in good agreement with the experimental value $1.7407 \pm 0.0059 \, GeV$
reported by the Particle Data Group. The number of expected $Z \to \nu
\bar\nu \gamma$ events for the considered sample is then:
\begin{gather}
  N^{(a)} = \frac{\Gamma^{(a)}}{\Gamma_{Z \to \textrm{had}}} 
  N_{Z \to \textrm{had}} =
  \frac{\alpha I}{3 \pi s_w^2 c_w^2 C} \, N_{Z \to \textrm{had}} \,
  \epsilon_6^2 \approx 54 \, \epsilon_6^2, \label{eq25}
  \\
  N^{(b)} = \frac{\Gamma^{(b)}}{\Gamma_{Z \to \textrm{had}}} \,
  N_{Z \to \textrm{had}} =
  \frac{8 M_Z^2 J}{3 \pi^2 v^2 C} \, N_{Z \to \textrm{had}} \, \epsilon_8^2
  \approx 52 \, \epsilon_8^2. \label{eq30}
\end{gather}

According to ref.~\cite{L3}, the number of \emph{background} events expected
from standard model is $\sim 2.4$, and the number of events experimentally
seen is $2$. So ordinary standard model background events completely cover
any possible new physics signal, and we can use experimental data only to
set an upper bound to the quantities~(\ref{eq25}) and~(\ref{eq30}). A rough
but simple way to do this is to require the expected signal to be smaller
than the observed background; in this way, we obtain a constraint for
$\epsilon_6$ and $\epsilon_8$:
\begin{gather}
  N^{(a)} < 2 \quad \Rightarrow \quad \epsilon_6 < 0.2, \label{eq40}\\
  N^{(b)} < 2 \quad \Rightarrow \quad \epsilon_8 < 0.2.
\end{gather}

In expressions (11) and (12) of ref.~\cite{Maya98}, Maya et al. report
constraints which is about 4 order magnitude smaller. It is not clear how
they managed to obtain such small values for $\epsilon_6$ and $\epsilon_8$.

The constraint~(\ref{eq40}) can be translated into an upper bound for
the $\tau$-neutrino magnetic monent, by means of eq.~(\ref{eq16}). In terms
of Bohr magneton, we have:
\begin{equation} \label{eq50}
  \mu_\nu = \frac{\epsilon_6}{\sqrt2 v} \frac{\mu_B}{\mu_B} =
  \frac{\epsilon_6}{\sqrt2 v \frac{e}{2 m_e}} \mu_B = 
  \frac{m_e}v \, \frac{\sqrt2 \epsilon_6}e \, \mu_B.
\end{equation}
The presence of the factor $\frac{m_e}v$ depends essentially on the fact that
we have chosen to measure $\mu_\nu$ in units of Bohr magnetons - which is a
quantity strictly related to electron mass - while $\mu_\nu$ has completely
nothing to do with electron properties. So our conclusion is that the factor
$\frac{m_e}v$ is not related to any mass scale involved in the calculation
of $\mu_\nu$ by means of $SU(2)_L \times U(1)_Y$ invariant quantities (as
Maya et al. claimed). This is also clear if we perform numerical substitution
in eq.~(\ref{eq50}) to extract an explicit result: if we assume for
$\epsilon_6$ the upper bound of $8.8 \times 10^{-5}$ trusted by Maya el al.,
their result $\mu_\nu < 1.14 \times 10^{-9}$ is reproduced, but if we use
our constraint~(\ref{eq40}) we obtain:
\begin{equation}
  \mu_\nu < 2 \times 10^{-6},
\end{equation}
which nicely coincide with the upper bound found by the L3 collaboration and
reported in~\cite{L3}. 

\section{Conclusions}

We have tried to understand the result announced by Maya et al. in
ref.~\cite{Maya98} for the $\tau$-neutrino magnetic moment, and we found
that their result is wrong. Moreover, we conclude that the use of $SU(2)_L
\times U(1)_Y$ gauge invariant operators do not improve bound on
$\mu_{\nu_\tau}$, in contrast to what Maya et al. claimed. Our calculations
reporduce and confirm the previous bound found by the L3 collaboration.

Investigation of M.~V. was supported by RFBR grant 98-02-17372.

\end{document}